\documentclass{Interspeech2024}

\usepackage{amsmath,graphicx}
\usepackage[utf8]{inputenc} % allow utf-8 input
\usepackage[T1]{fontenc}    % use 8-bit T1 fonts
\usepackage{hyperref}       % hyperlinks
\usepackage{url}            % simple URL typesetting
\usepackage{makecell}       % professional-quality tables
\usepackage{booktabs}
\usepackage{amsfonts}       % blackboard math symbols
\usepackage{nicefrac}       % compact symbols for 1/2, etc.
\usepackage{microtype}      % microtypography
\usepackage{xcolor}         % colors
\usepackage{bm}
\usepackage{wrapfig}
\usepackage{caption}
\usepackage{algorithm}
\usepackage{algorithmic}
\usepackage{bm}
\usepackage{graphicx}
\usepackage{amsmath}
\usepackage{amssymb}
\usepackage{mathtools}
\usepackage{amsthm}
\usepackage{multirow}
\usepackage{tabularx}
\usepackage{subcaption}
\usepackage{tcolorbox}
\usepackage{enumitem}
\usepackage[formats]{listings} % code formatting package
\usepackage{lscape} % horizontal 

\interspeechcameraready 

\title{Exploring In-Context Learning of Textless Speech Language Model for Speech Classification Tasks\thanks{*The first two authors contributed equally.}}

\name[affiliation={*,1}]{Ming-Hao}{Hsu}
\name[affiliation={*,2}]{Kai-Wei}{Chang}
\name[affiliation={3}]{Shang-Wen}{Li}
\name[affiliation={1}]{Hung-yi}{Lee}

\address{
  $^1$Department of Electrical Engineering, National Taiwan University, Taiwan \\
  $^2$Graduate Institute of Communication Engineering, National Taiwan University, Taiwan\\
  $^3$Meta AI, USA}
\email{qaz159qaz159@gmail.com, kaiwei.chang.tw@gmail.com}

\keywords{In-context learning, speech language model, prompt tuning, few-shot learning, speech classification}

\begin{document}

\maketitle

\begin{abstract}
Ever since the development of GPT-3 in the natural language processing (NLP) field, in-context learning (ICL) has played an essential role in utilizing large language models (LLMs).
By presenting the LM utterance-label demonstrations at the input, the LM can accomplish few-shot learning without relying on gradient descent or requiring explicit modification of its parameters.
This enables the LM to perform various downstream tasks in a black-box manner. 
Despite the success of ICL in NLP, little work is exploring the possibility of ICL in speech processing.
This study is the first work exploring ICL for speech classification tasks with textless speech LM.
We first show that the current speech LM lacks the ICL capability.
We then perform warmup training on the speech LM, equipping the LM with demonstration learning capability. 
This paper explores and proposes the first speech LM capable of performing unseen classification tasks in an ICL manner.
\end{abstract}
%
% \begin{keywords}
% In-context learning, speech language model, prompt tuning, few-shot learning, speech classification
% \end{keywords}
%
\section{Introduction}
Large language models (LLMs)~\cite{zhao2023survey, bommasani2021opportunities} have gained significant attention in recent years. 
With the development of LLMs like GPT-3~\cite{brown2020language}, researchers have discovered the potential for performing \textbf{in-context learning (ICL)}~\cite{brown2020language, dong2022survey}.
ICL is a technique that enables LMs to learn to perform new tasks from a small number of demonstrations which are presented at the input of the LM.
Formally, we consider a set of data points, denoted as $x_i$, along with their corresponding labels, denoted as $y_i$~\footnote{
In this paper, we consider ICL following the definition in \cite{dong2022survey}. That is, the model learns the relationship between the ``data point $x$'' and its ``label $y$'' by providing them as demonstrations. Recent works, such as VALL-E~\cite{valle} and AudioBox~\cite{audiobox}, also claim to perform ICL with large speech models. However, they primarily focus on learning the acoustic condition or speaker identity of the given speech prompt. This topic has been studied in the speech processing literature as ``voice cloning~\cite{arik2018neural}'' and ``style transfer~\cite{wang2018style}'' and is outside the scope of this paper.}. 
Additionally, we have a target data point, $x_t$, for which we want the LM to make an inference. To achieve this, we prepend the demonstrations, consisting of the data points and their labels, to the input sequence as follows: $[x_1, y_1, x_2, y_2, ..., x_n, y_n, x_t]$. By learning the analogy between the data points and their labels, the LM is capable of directly predicting the label of $x_t$. It is important to note that throughout this learning process, the LM remains fixed, and there is no gradient backward process involved. Instead, the LM relies solely on the input demonstrations to acquire knowledge.

ICL, as an \emph{emergent ability}~\cite{wei2022emergent, webb2022emergent} of the LLM, remains insufficiently optimized due to its disparity between the LM's pre-training task. Regarding this, researchers have attempted to perform \emph{warmup training}~\cite{dong2022survey} in either a supervised~\cite{min2022metaicl} or self-supervised~\cite{chen2022improving, DBLP:conf/acl/Gu0WH23} manner to enhance the ICL capability. 
The motivation for building an LM with ICL capability is that, as a new paradigm, ICL offers several advantages.
First, ICL simplifies the integration of human knowledge into the LM by providing demonstrations. This process resembles the analogy reasoning process of humans~\cite{zhao2023survey}.
Second, ICL incorporates demonstrations at the input, eliminating the need for backpropagation and gradient flow to establish connections between data points and their labels. As a result, computational costs are reduced.
Third, LLMs are often released as a service in the real-world application~\cite{brown2020language, sun2022black, chatgpt}. ICL is particularly suitable for LM deployment since it only modifies the input, allowing LLMs to adapt to new tasks defined by the users~\cite{sun2022black}.
Overall, in the NLP field, ICL has emerged as a powerful paradigm for utilizing LLMs. However, despite the recent advancements in large speech LMs, there is a notable lack of research on ICL in the domain of speech processing. Fortunately, the recently developed \textbf{textless speech language models (speech LMs)} offer an opportunity for us to explore ICL for speech processing.

\begin{figure}[t]
    \centering
\includegraphics[width=\linewidth]{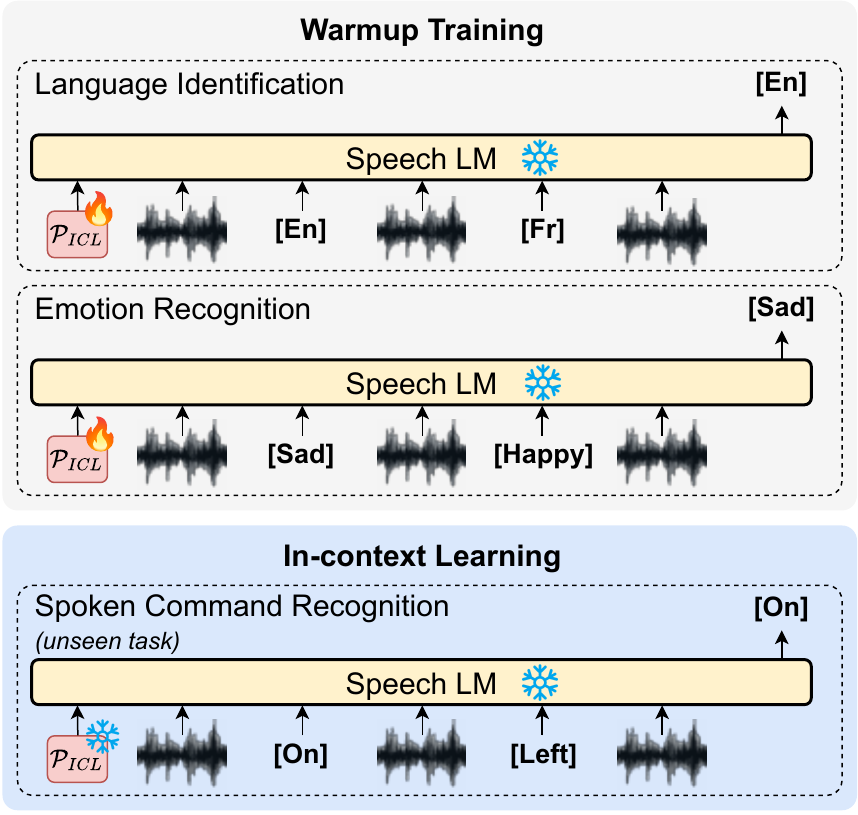}
  \caption{This figure illustrates the framework of the proposed approach, where warmup training utilizes a variety of training tasks to instill in-context learning abilities in the speech language model. This allows the model to utilize task demonstrations effectively to tackle novel unseen tasks.}
    \label{fig:ICL}
\end{figure}
% In recent years, textless \textbf{speech language models (speech LMs)} have emerged. 
Textless speech LMs quantize speech representations into \emph{discrete speech tokens}~\cite{DBLP:conf/interspeech/PolyakACKLHMD21} and engage in the next token prediction pre-training task, akin to language models in the NLP field. These textless speech LMs are trained on large speech-only datasets and demonstrate robust generation capabilities without any text supervision.
Demonstrating robust capabilities, these speech LMs can generate meaningful discrete tokens conditioned on given speech segments. 
Notable examples include the Generative Spoken Language Model (GSLM)\cite{lakhotia2021generative}, pGSLM\cite{kharitonov2022text},  audioLM~\cite{borsos2023audiolm} and others.
Although recently there have been researches utilizing text data to help speech language modeling (e.g. TWIST~\cite{hassid2023textually}, VALL-E~\cite{valle}), we focus on textless speech LM in this work. The reason is that the textless property offers appealing advantages, including but not limited to:
(1) \textbf{Language agnostic}: When performing downstream tasks, textless speech LM is not restricted to text prompts with a specific language~\cite{valle}. (2) \textbf{Direct speech modeling}: Textless speech LM directly models the speech signal, which bypasses the need for speech-text paired data that is usually expensive to collect. Furthermore, many languages lack a written or formal textual form. Overall, the textless property offers us a more general and easy-to-scale framework.

We first examined the ICL ability of the current largest open-sourced textless speech LM, GSLM. We observed that GSLM failed to comprehend the provided speech-label pairs for speech classification tasks, indicating the lack of emergent ability to perform ICL.
If the LM is capable of performing ICL, it makes predictions by learning the input-label mapping~\cite{wei2023larger}, even with random labels~\cite{min2022rethinking}. A possible explanation is the LM performs ``implicit fine-tuning'' during ICL~\cite{dai2023can}. The preliminary result shows the current speech LM does not possess such emergent ability~\footnote{In fact, we also examined a more powerful speech LM, TWIST~\cite{hassid2023textually}, that adopts text data for pre-training. However, we do not observe the ICL capability either.}.
As shown in Fig.~\ref{fig:ICL}, to build a speech LM with ICL ability, we conduct a simple warmup training with parameter efficient tuning (PET) method, prompt tuning~\cite{liu2023pre, chang22e_interspeech}. Warmup training is performed on a set of training tasks to enable the speech LM to understand the demonstrations and make predictions when encountering unseen tasks.

The experimental results indicate that GSLM, when subjected to warmup training, demonstrates the capability to perform ICL not only on seen tasks but also, surprisingly, on unseen tasks. It surpasses the random guessing baseline for all tasks and can outperform the support vector classifier (SVC) in most scenarios.
It's worth noting that, in this paper, we aim to show the feasibility of ICL for speech LM, not outperform the current state-of-the-art methods. We believe that as more advanced textless speech LM continues to be developed, the significance of ICL behavior will become increasingly evident, mirroring the observations made in the NLP field (a notable example is the evolution from GPT-2 to GPT-3~\cite{min2022rethinking}).
Our contributions are as follows:
% \begin{itemize}[itemsep=-0.4em,leftmargin=*]
\begin{itemize}
    \item We investigate the in-context learning capability of the existing speech LM and identify its limitations in this regard.
    \item We introduce the first speech LM that incorporates warmup training, enabling it to perform in-context learning effectively. This is the first speech LM with such capabilities.
    \item We empirically demonstrate that the speech LM can effectively learn and adapt to unseen tasks through ICL and achieve non-trivial results, surpassing the performance of a random sampling baseline.
\end{itemize}
\begin{table*}[t]
    \centering
    \caption{This table displays the accuracy for ICL with warmup training (w/ Warmup), random guessing (Random), and ICL without warmup training (w/o Warmup) and SVC on seen tasks ($\mathcal{T}_{train}$) and unseen tasks ($\mathcal{T}_{test}$) for the two task groups. The results demonstrate that warmup training enables effective in-context learning, outperforming baselines on diverse speech tasks.}
    \begin{tabular}{@{}c|c|c|c|>{\centering\arraybackslash}p{1.7cm}|>{\centering\arraybackslash}p{1.7cm}|>{\centering\arraybackslash}p{1.7cm}|>{\centering\arraybackslash}p{1.7cm}>{\centering\arraybackslash}p{1.7cm}@{}}%\toprule
    \Xhline{2pt}
    Group & Task Type & Task & Dataset & w/ Warmup & Random & w/o Warmup & SVC \\
    % \cmidrule{1-8}
    \Xcline{1-8}{0.4pt}
    \multirow{8}*{Group 1}  & \multirow{2}*{$\mathcal{T}_{test}$} & SCR & Arabic SC ~\cite{benamer2020database} & 40.9 \textsubscript{$\pm$ 1.5} & 28.0 \textsubscript{$\pm$ 1.0} & 3.9 \textsubscript{$\pm$ 0.7} & \textbf{50.8}\\

    && ER  & IEMOCAP ~\cite{busso2008iemocap} & \textbf{47.7} \textsubscript{$\pm$ 1.8} & 41.0 \textsubscript{$\pm$ 1.4} & 3.1 \textsubscript{$\pm$ 0.4} & 33.7\\
    % \cmidrule{2-8}
    \Xcline{2-8}{0.4pt}
      & \multirow{5}*{$\mathcal{T}_{train}$} & SCR & Google SC v2 ~\cite{warden2018speech} & \textbf{79.6} \textsubscript{$\pm$ 1.1} & 25.1 \textsubscript{$\pm$ 1.6} & 6.4 \textsubscript{$\pm$ 0.8} & 43.8\\

                          & & SCR & Lithuanian SC ~\cite{kolesau2020unsupervised} & \textbf{80.5} \textsubscript{$\pm$ 1.1} & 25.1 \textsubscript{$\pm$ 1.6} & 8.4 \textsubscript{$\pm$ 1.2} & 37.8\\

                           && SCR & Dysarthric Mandarin SC ~\cite{app11062477} & \textbf{57.8} \textsubscript{$\pm$ 0.8} & 24.7 \textsubscript{$\pm$ 0.9} & 9.8 \textsubscript{$\pm$ .0.9} & 12.0\\

                           && LID & Voxforge ~\cite{maclean2018voxforge} & \textbf{32.0} \textsubscript{$\pm$ 1.6} & 23.8 \textsubscript{$\pm$ 2.7} & 1.8 \textsubscript{$\pm$ 0.5} & 29.7\\

                           && SD  & MUStARD ~\cite{DBLP:conf/acl/CastroHPZMP19} & \textbf{64.7} \textsubscript{$\pm$ 1.6} & 54.7 \textsubscript{$\pm$ 1.0} & 1.2 \textsubscript{$\pm$ 0.2} & 60.9\\
    % \cmidrule{1-8}
    \hline
    \hline
    \multirow{8}*{Group 2}  & \multirow{3}*{$\mathcal{T}_{test}$} & SCR & Arabic SC ~\cite{benamer2020database} & 36.5 \textsubscript{$\pm$ 1.2} & 28.0 \textsubscript{$\pm$ 1.0} & 3.9 \textsubscript{$\pm$ 0.7} & \textbf{50.8}\\

          &                & SCR & Google SC v2 ~\cite{warden2018speech} & \textbf{48.0} \textsubscript{$\pm$ 0.7} & 25.1 \textsubscript{$\pm$ 1.6} & 6.4 \textsubscript{$\pm$ 0.8} & 43.8\\ 

           &                & SD & MUStARD ~\cite{DBLP:conf/acl/CastroHPZMP19} & \textbf{64.1} \textsubscript{$\pm$ 1.1} & 54.7 \textsubscript{$\pm$ 1.0} & 1.2 \textsubscript{$\pm$ 0.2} & 60.9\\
    \Xcline{2-8}{0.4pt}
                        & \multirow{4}*{$\mathcal{T}_{train}$} & SCR & Dysarthric Mandarin SC ~\cite{app11062477} & \textbf{56.0} \textsubscript{$\pm$ 1.9} & 24.7 \textsubscript{$\pm$ 0.9} & 9.8 \textsubscript{$\pm$ 0.9} & 12.0\\

                         &  & SCR & Lithuanian SC ~\cite{kolesau2020unsupervised} & \textbf{80.5} \textsubscript{$\pm$ 0.9} & 25.1 \textsubscript{$\pm$ 1.6} & 8.4 \textsubscript{$\pm$ 1.2} & 37.8\\

                          & & LID & Voxforge ~\cite{maclean2018voxforge} & \textbf{29.9} \textsubscript{$\pm$ 0.5} & 23.8 \textsubscript{$\pm$ 2.7} & 1.8 \textsubscript{$\pm$ 0.5} & 29.7\\

                           && ER & IEMOCAP ~\cite{busso2008iemocap} & \textbf{47.6} \textsubscript{$\pm$ 1.7} & 41.0 \textsubscript{$\pm$ 1.4} & 3.1 \textsubscript{$\pm$ 0.4} & 33.7\\
    % \cmidrule{1-8}
    \Xhline{2pt}
    \end{tabular}
    \label{tab:main_table}
\end{table*}

\section{Method}
\label{sec:Method}

We first investigated the ICL ability of the pre-trained GSLM~\cite{lakhotia2021generative}, which is the largest open-sourced textless generative speech LM.
In this paper, we focus on speech classification tasks. 
We provide speech-label pairs as demonstrations at the input and let the model predict the label of a target utterance.
% ~\footnote{Note that in speech classification tasks, it's essential to establish a mapping process from labels to LM's vocabulary. This allows the speech LM to generate labels using its vocabulary, with consideration given to discrete speech tokens.
% Previous studies~\cite{min2022rethinking} have shown that random label mapping is feasible for the LM to conduct ICL. And the LM might perform ``implicit fine-tuning'' during ICL. Without loss of generalizability, we also conduct random label mapping to study the ICL of the speech LM.}. 
Our findings indicate that the current GSLM does not possess ICL ability. As shown in Table~\ref{tab:main_table}, ``w/o Warmup'' is the performance of directly applying ICL on GSLM without warmup training. ``Random'' is the performance of randomly guessing a label for the speech classification tasks. We find that applying ICL directly to GSLM yields results worse than random guessing.

To address this limitation and build a speech LM with ICL ability, we propose conducting \textbf{warmup training} on the speech LM. For the warmup training, we utilize a set of training tasks denoted as $\mathcal{T}_{train}$ to enhance the speech LM's ICL capability. Specifically, we employ a PET method, prompt tuning~\cite{liu2023pre, chang22e_interspeech, chang2023prompting}. Applying prompt tuning is a design choice, and other tuning methods can also be adopted in the warmup training, as discussed in \cite{dong2022survey}. 
However, We conduct prompt tuning for two reasons: (1) Preliminary experiments revealed that fine-tuning the entire model for warmup training is unstable and might lead to inferior ICL performance. (2) Prompt tuning prepends prompt vectors at the input side while keeping the pre-trained speech LM fixed. This preserves its generative capability, which is beneficial for future applications.

In the following sections, we outline our approach to conducting warmup training on the set of training tasks $\mathcal{T}_{train}$ and evaluate the ICL capability of the model on the training tasks $\mathcal{T}_{train}$ (seen tasks) and testing tasks $\mathcal{T}_{test}$ (unseen tasks).
\subsection{Warmup Training}
Given a speech LM $\mathcal{M}$ that performs next token prediction on discrete speech tokens $x$:
\begin{equation}
    x_{t+1} = \mathcal{M}(x_1, x_2, ...,x_t),
\end{equation}
where $t$ is the timestep. We first collect a set of training tasks $\mathcal{T}_{train}$ to perform warmup training.
Each task $T_i$ in $\mathcal{T}_{train}$ uses its own dataset.
To form one training data point for ICL warmup training, we conduct the following procedure:

(1) We randomly sample $n$ utterances and their corresponding labels from a training task as demonstrations.
(2) Following GSLM~\cite{lakhotia2021generative}, the utterances are first encoded into discrete token sequences $\bm{x}_{1}, \bm{x}_{2}, ..., \bm{x}_{n}$.
In addition, we utilize \textbf{random label mapping (RLM)}~\cite{chen2023understanding, tsai2020transfer, min2022rethinking} to map the task's labels (target domain) to the discrete tokens (source domain), resulting in labels for the demonstrations $y_1, y_2, ..., y_n$.
(3) Each speech token sequence is then truncated or padded to the same utterance length $L$, yielding $\tilde{\bm{x}}_{1}, \tilde{\bm{x}}_{2}, ..., \tilde{\bm{x}}_{n}$. We found this step critical as it provides a standardized format to the speech LM and simplifies the training.
(4) The input data is constructed as 
\begin{equation}
    X = [\tilde{\bm{x}}_{1}, \langle s \rangle, y_{1}, \langle s \rangle, ..., \tilde{\bm{x}}_{n}, \langle s \rangle, y_{n}, \langle s \rangle, \tilde{\bm{x}}_{t}, \langle s \rangle],
    \label{equ:demon}
\end{equation}
where ``$\langle s \rangle $'' is a separation token with trainable embedding, and $\tilde{\bm{x}}_{t}$ is the target utterance, which is the data we want the model to predict its label.

During warmup training, we randomly sample the target utterance from the demonstrations, that is, $\tilde{\bm{x}}_{t} \in \{\tilde{\bm{x}}_{1}, \tilde{\bm{x}}_{2}, ..., \tilde{\bm{x}}_{n}\}$.
The model then learns to compare the target utterance and the demonstrations in order to predict the correct label.
We find this step simple but effective since it simplifies the training objective. The model is required to compare the target utterance with each demonstration and output its corresponding label. The learned behavior benefits ICL in the next stage.
We then employ prompt tuning to learn the prompts $\mathcal{P}_{ICL}$ to equip the model with ICL capability. Following ~\cite{chang2023speechprompt}, a set of prompt vectors $\mathcal{P}_{ICL}$ are prepended at the input of the speech LM. The speech LM then makes the prediction conditioned on the demonstrations $X$ and the prompt vectors $\mathcal{P}_{ICL}$. We then apply cross entropy loss $\mathcal{L}$ on the model prediction and the ground truth label of the target utterance $y_t$ for optimizing the prompts:
\begin{equation}
    \mathcal{P}_{ICL} \leftarrow \mathop{\arg\min}_{\mathcal{P}_{ICL}} \mathcal{L}(\mathcal{M}(X; \mathcal{P}_{ICL}), y_t)),
\end{equation}
% \begin{equation}
%     \mathcal{L} = CE(\mathcal{M}(X; \mathcal{P}_{ICL}), y_t)),
% \end{equation}
where $\mathcal{M}(X; \mathcal{P}_{ICL})$ represents the model prediction conditioned on the input data $X$ and the prompts $\mathcal{P}$, and $y_t$ is the ground truth label corresponding to the target utterance $\tilde{\bm{x}}_{t}$.

\subsection{In-context Learning}
After completing the warmup training, the model becomes ready to perform ICL on the training tasks $\mathcal{T}_{train}$ (seen tasks) and testing tasks $\mathcal{T}_{test}$ (unseen tasks). The process of preparing demonstrations during this stage is similar to the warmup stage as described in the formula (\ref{equ:demon}). 
The difference is that, in the ICL stage, the target utterance $\tilde{\bm{x}}_{t}$ is no longer included in the demonstrations. Instead, its corresponding label ${y}_{t} \in \{{y}_{1}, {y}_{2}, ..., {y}_{n}\}$ is included, enabling the model to learn to make predictions based on analogies.

\section{Experimental Setup}
\label{sec:experimentalsetup}
\subsection{Tasks and Datasets}
We evaluate speech LM's ICL ability on a diverse set of speech classification tasks with 7 datasets. These tasks include speech command recognition (SCR), emotion recognition (ER), language identification (LID), and sarcasm detection (SD). Also, the datasets involve varying languages, accents, domains, and label distributions, allowing a comprehensive evaluation of the transferability of ICL for speech LM.
As shown in Table \ref{tab:main_table}, we select two groups (Group 1 and Group 2) of training tasks $\mathcal{T}_{train}$ and testing tasks $\mathcal{T}_{test}$. These tasks are combined in ways that introduce variety, for instance, tasks with different numbers of classes and different types of tasks. This approach enables us to evaluate our model's performance in ICL and gauge the impact of the training tasks on performance across a range of testing tasks. Our attention is directed explicitly toward the model's capacity to utilize learned ICL knowledge during warmup training when faced with a new task. This capability is essential for many practical uses in a multitude of real-world situations.

We've chosen particular datasets for both training and testing. Our aim for each training dataset group includes generating a balanced dataset to avoid bias. To meet this goal, we sample 10,000 data points, each including 4 demonstrations from the training tasks. The same amount of data points is ensured for every task, providing a balanced dataset as each task gets an equal proportion of demonstrations. If we neglect to do this and simply use the entire dataset, some of the larger datasets would dominate a significant portion of the combined dataset, resulting in a skewed dataset.

\subsection{Implementation Detail}
We adopt GSLM~\cite{lakhotia2021generative} as our backbone speech LM. Specifically, the GSLM is trained on top of discrete units encoded by HuBERT~\cite{hsu2021hubert} SSL speech model and K-means clustering algorithm with 100 clusters.
In warmup training, we conduct prompt tuning and use a prompt length equal to 5. This approach introduces a small fraction of trainable parameters, specifically less than 0.1\% of the total 150 million parameters of the speech LM, simplifying the learning process.
As described in Section~\ref{sec:Method}, our initial approach involves enforcing a fixed length for utterances. In the primary experiment in Sec.~\ref{sec:main_result}, we fix the utterance length $L$ to be 50. This standardization ensures consistent utterance lengths across multiple datasets and simplifies the training. We also investigate the impact of varying utterance lengths in Sec.~\ref{subsec:utterance_length}.
We provide four demonstrations and one target utterance in both warmup training and the ICL stage~\footnote{In our preliminary study, we find four demonstrations as a suitable setup, for it provides a reasonable demonstration number while preventing the ``curse of long sequence'' problem as discussed in the previous work~\cite{chang2023speechprompt}. How to incorporate more demonstrations and alleviate the long-form problem remains future work.}.

Given the limited research of ICL on speech LM, we compare our GSLM with warmup training (\textbf{w/ Warmup}) with three baselines: (1) random guessing (\textbf{Random}), (2) ICL on GSLM without warmup training (\textbf{w/o Warmup}), and (3) a support vector classifier (\textbf{SVC}).
The random guessing method entails making predictions by selecting labels at random from the demonstrations; while SVC is trained using the provided demonstrations to make predictions. We repeat the experiments five times and compute the average accuracy alongside its standard deviation, offering a more fair evaluation of the model's performance.

\section{Results}
\label{sec:results}

\subsection{Main Result}
\label{sec:main_result}
% In Table~\ref{tab:main_table}, we provide a detailed performance comparison of our proposed ICL with warmup training (\textbf{w/ Warmup}), random guessing (\textbf{Random}), ICL without warmup training (\textbf{w/o Warmup}), and a support vector classifier (\textbf{SVC}). This performance comparison is given across two different task groups (Group 1 and Group 2) for both training tasks ($\mathcal{T}_{train}$) and testing tasks ($\mathcal{T}_{test}$).

In Table~\ref{tab:main_table}, the result shows that ICL with warmup training consistently outperforms both the Random and w/o Warmup baselines and surpasses SVC for most tasks. For Group 1, in unseen tasks $\mathcal{T}_{test}$, ICL with warmup training excels on the Arabic SC dataset with a score of 40.9, notably higher than both Random and w/o Warmup, but lower than SVC's 50.8. In the IEMOCAP dataset, w/ Warmup surpasses all baselines, including SVC.
In seen tasks $\mathcal{T}_{train}$, the w/ Warmup method is superior, particularly scoring 79.6 on Google SC v2 and 80.5 on Lithuanian SC, outpacing all baselines.
Group 2 follows a similar trend; the w/ Warmup method leads across the board. In test tasks on Google SC v2, it scores 48.0, markedly higher than both Random and w/o Warmup and slightly better than SVC. 

Overall, in both Group 1 and Group 2, ICL with Warmup outperforms other baselines. The only exception is the Arabic SC task presented as an unseen task. Although GSLM can perform ICL when such cross-lingual tasks are in $\mathcal{T}_{train}$ (such as Lithuanian SC and Mandarin SC), we hypothesize that for GSLM, it's still a challenge when presenting cross-lingual tasks as unseen tasks $\mathcal{T}_{test}$.
These results underline the efficacy of the warmup training for ICL, as it consistently outperforms both baselines across a diverse range of speech tasks. This success opens up new paths for future studies and holds potential for more improvements in this field.

\label{sec:ablation}
\subsection{Model Behavior Analysis}
\begin{table}[]
    \centering
    \caption{Guessing Rate Analysis. The results demonstrate that warmup training significantly improves the model's ability to make predictions based on the provided demonstrations in ICL.}
    \resizebox{\linewidth}{!}
    {
    \begin{tabular}{@{}c|c|c|c@{}}%\toprule
    \Xhline{2pt}
    \multirow{2}{*}{Group} & \multirow{2}{*}{Task Type} & \multicolumn{2}{c}{Guessing Rate (Avg)}\\
    % \cmidrule{3-4}
    \Xcline{3-4}{0.4pt}
    & & w/o Warmup & w/ Warmup\\
    % \cmidrule{1-4}
    \Xcline{1-4}{0.4pt}
    \multirow{2}*{Group 1}&$\mathcal{T}_{test}$ & 13.9 & 95.1\\
    &$\mathcal{T}_{train}$ & 21.0 & 98.4\\
    % \cmidrule{1-4}
    \Xcline{1-4}{0.4pt}
    \multirow{2}*{Group 2}&$\mathcal{T}_{test}$ & 15.0 & 92.2\\
    &$\mathcal{T}_{train}$ & 21.7 & 97.4\\
    % \cmidrule{1-4}
    \Xhline{2pt}
    \end{tabular}
    }
    \label{tab:guess_rate}
\end{table}

Warmup training aims to equip the model with the ability to identify, compare demonstrations, and comprehend the target task. 
We assess the model's ability to predict labels based on demonstrations by examining its guessing rate during ICL.
The guessing rate indicates the probability of the model identifying demonstrated labels. Higher guessing rates demonstrate effective ICL, with the model making predictions derived from the demonstrations.
Table~\ref{tab:guess_rate} compares guessing rates between models with and without warmup training across two groups, including seen and unseen tasks. Models lacking warmup training exhibit low guessing rates in both Group 1 and Group 2. Conversely, warmup training significantly boosts guessing rates to over 90\%.

\begin{figure}
    \centering
    \includegraphics[width=\linewidth]{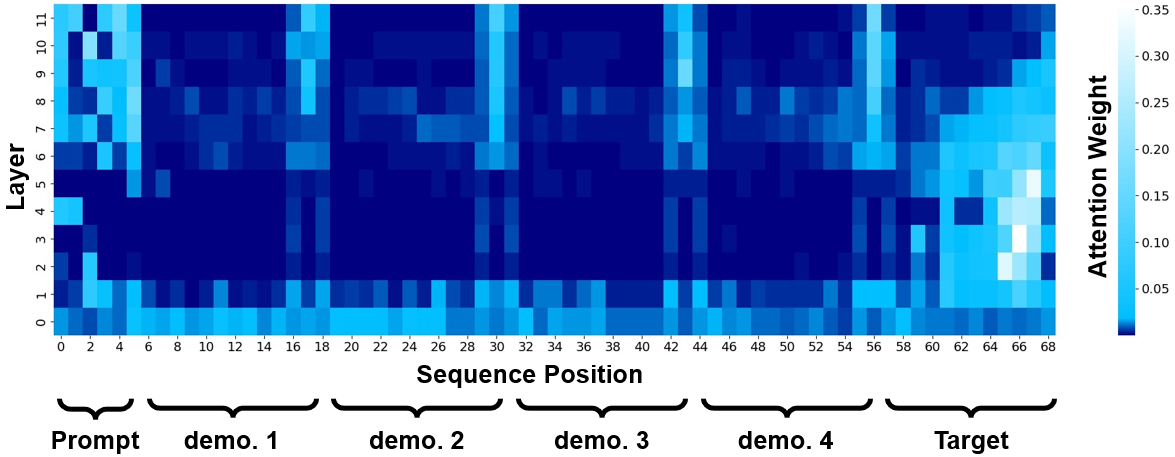}
    \caption{Model Predicting Attention Weights in ICL.
    Initially, the model primarily attends to the demonstrations within the first two layers. It then gradually shifts its focus to the target in layers 3 through 7, and finally towards the labels in the demonstrations in the final layers. For the demonstration purpose, we show the scenario where the utterance length $L$ is 10.}
    \label{fig:attention}
\end{figure}

We further examine the model's behavior while predicting the label in ICL.
The attention map in the model's attention layers during the execution of ICL is depicted in Figure~\ref{fig:attention}. 
The figure reveals that the initial two layers mainly concentrate on the demonstrations. 
The focus then shifts to the target utterance in the middle layers (3rd to 7th) and finally shifts to the demonstrations' labels in the last layers (8th to 12th). 
Also, the model's continued attention to the prompts $\mathcal{P}_{ICL}$ highlights the utility of warmup training in ICL. 
From the studies in this section, we can see that the warmup training effectively steers the model to perform ICL for unseen tasks.

\subsection{Utterance Length Analysis}
\label{subsec:utterance_length}
\begin{table}[]
    \centering
    % \caption{Utterance Length Analysis. Accuracy comparison across different utterance lengths. An appropriate utterance length (e.g. 30-50 units) provides the best performance by balancing information and differentiation between examples.}
    \caption{Utterance Length Analysis. Accuracy comparison across different discrete tokens in the utterance is reported. }
    \resizebox{\linewidth}{!}
    {
    \begin{tabular}{@{}cccccc@{}}%\toprule
    \Xhline{2pt}
    \multicolumn{2}{c}{\textbf{\# of Tokens}}& 10 &30 &50 &Not Fixed \\
    % \cmidrule{1-5}
    \Xhline{0.4pt}
    \multirow{2}{1.2cm}{\centering Group 1 $\mathcal{T}_{test}$}&Arabic SC & 40.2 & \textbf{41.7} & 40.9 & 23.7 \\
    &IEMOCAP & 42.2 & 45.7 & \textbf{47.7} & 44.8 \\
    % \cmidrule{1-5}
    \Xhline{2pt}
    \end{tabular}
    }
    \label{tab:length_table}
\end{table}

As shown in Table \ref{tab:length_table}, the model's performance is influenced by the number of discrete tokens in each utterance. If the length of the utterance is too long, although it might hold sufficient information, the GSLM could struggle with modeling such long sequences as reported in ~\cite{chang22e_interspeech}.
On the other hand, if utterances are too short, the model may lack the necessary information, leading to random guessing. We expect with more advanced speech LM built, more demonstrations with longer length can be incorporated to boost the performance.

% \subsection{Prompt Length Analysis}
% \input{tables/prompt_length_analysis}
% Table \ref{tab:prompt_table} shows that overly long prompt lengths can lead to reduced overall performance. This situation shows that an excessive number of parameters could cause overfitting on the training datasets. Therefore, using too many prompts (like 100) will lead to lower performance than using fewer prompts.
\section{Conclusion}
\label{sec:conclusion}

This paper presents the first successful application of in-context learning (ICL) to speech LM. We initially investigated the limitations of the current speech LM in performing ICL. With the proposed warmup training, the textless GSLM demonstrates ICL capability on seen tasks and successfully achieves non-trivial results on unseen tasks across diverse datasets.
This paper does not aim to achieve competitive performance with ICL but to show its feasibility. We're also aware that the current capacity of the backbone GSLM is restricted. GPT-3 contains 175B parameters, while GSLM is 1000 times smaller, containing 150M parameters, which might cause the limitation of the ICL emergent ability. Future works include investigating ICL on more diverse speech LM and developing more effective warmup strategies for ICL.
\section{Acknowledgement}
\label{sec:acknowledgement}
We thank the National Center for High-performance Computing (NCHC) of National Applied Research Laboratories (NARLabs) in Taiwan for providing computational and storage resources.

\bibliographystyle{IEEEtran}
\bibliography{refs}

\end{document}